\documentclass[fleqn,10pt]{wlscirep}
\usepackage{amsmath}
\usepackage{graphicx}
\usepackage{footmisc}
\usepackage{color}

\begin{document}
\title {Persistent current in 2D topological superconductors}
\author[1]{Igor N. Karnaukhov}
\affil[1]{G.V. Kurdyumov Institute for Metal Physics, 36 Vernadsky Boulevard, 03142 Kiev, Ukraine}
\affil[*]{karnaui@yahoo.com}
\begin{abstract}
A junction between two boundaries of a topological superconductor (TSC), mediated by localized edge modes of Majorana fermions, is investigated. The tunneling of fermions across the junction depends on the magnetic flux. It breaks the time-reversal symmetry at the boundary of the sample. The persistent current is determined by the emergence of Majorana edge modes. The structure of the edge modes depends on the magnitude of the tunneling amplitude across the junction. It is shown that there are two different regimes, which correspond to strong and weak tunneling of Majorana fermions, distinctive in the persistent current behavior. In a strong tunneling regime, the fermion parity of edge modes is not conserved and the persistent current is a $2\pi$-periodic function of the magnetic flux. When the tunneling is weak the chiral Majorana states, which are propagating along the edges have the same fermion parity. They form a $4\pi$-phase periodic persistent current along the boundaries. The regions in the space of parameters, which correspond to the emergence of $2\pi$- and of $4\pi$-harmonics, are numerically determined. The peculiarities in the persistent current behavior are studied.
\end{abstract}
\maketitle

\section*{Introduction}

The phase coherent tunneling across a junction between two superconductors implies the presence of a $2 \pi-$ periodic persistent current, which is defined by the phase difference between the superconducting order parameters. The Josephson effect has been considered in Refs \cite{I1,Er} in the framework of the well-known Kitaev chain model \cite{Kitaev}. The Kitaev's proposal that, in the case of the fermion parity conservation, zero-energy Majorana fermion states, which are localized at the ends of the superconducting wire, trigger a  $4 \pi$-periodic persistent current explains the so-called 'topological (or fractional) Josephson effect. The $2 \pi$- and $4 \pi$-harmonics of a persistent current correspond to the respective ground states of the system with different fermion parity when the magnetic flux is greater than $\pi$. In \cite{Kitaev} the author stimulates further research of new topological states that are realized at junctions between 1D TSCs, and Luttinger liquids \cite{I2,I3}. In the absence of fermion parity conservation (that is, in those superconductors, in which the total number of particles is not conserved), the system under consideration is relaxing to the phase state with the lowest energy, which leads to the emergence of a $2\pi$-periodic persistent current.

Below we discuss the persistent current in a 2D $(p+ip)$ TSC that has the spatial form of a hollow cylinder and is penetrated by a magnetic flux Q. We expect a nontrivial behavior of the persistent current depending on the magnitude of the applied magnetic flux. Due to their nontrivial topology \cite{TSC5,TSC6,TSC7,TSC8}, the superconductors with $(d +id)$ and $(p+ip)$ order parameters exhibit exotic phenomena such as Majorana vortex bound states and gapless chiral edge modes. The 2D TSCs with the ($p+ip)$-pairing of spinless fermions, which have chiral Majorana fermion states propagating along the edges, have been considered in  \cite{TSC5a}.  The behavior of topological states in the presence of disorder has been studied in Refs \cite{A1,D1,D3,D4,D5,D6}. A nontraditional approach for description of TSCs has been proposed in \cite{K1} (see also  \cite{K2}). It was shown that spontaneous breaking of time reversal symmetry is realized due to nontrivial stable phases of the superconducting order parameter (new order parameter). At that, the models of the TSC with the $p-$ and $(p+ip)$-wave superconducting pairing of spinless fermions are the simplest and the most straightforward examples of relevant model systems.

In a finite system, the gapless chiral edge modes are localized at the boundaries. The tunneling of fermions across a junction leads to gapped edge modes due to the hybridization (through the weak link) of chiral edge modes localized at the different boundaries of the junction. In the case of a 1D superconductor the fermion parity is associated with zero energy Majorana edge states \cite{Kitaev,I5,I6,I7,D7}, for a 2D TSC a persistent current is determined by the presence of the Majorana gapless edge modes localized at the boundaries of the junction. The ground state fermion parity changes whenever the energy of a pair of Majorana fermions crosses the zero energy. In the superconductor-topological insulator system the fermion parity of the ground state was associated with the Hopf index \cite{I4}. The fermion parity conservation, as a rule, is the result of the conservation of the total number of particles in the system, while the total number of particles is not conserved in those superconductors, which were studied in the framework of the Bogoliubov-de Gennes formalism. Nevertheless, we show that the fermion parity conservation is realized due to the conservation of the Chern number that determines the chiral current at the ends of the cylinder. The key point of the paper is that the unconventional behavior of the persistent current is determined by a chiral current along the boundaries of the TSC, while the behavior of the persistent current depends on the value of the tunneling amplitude of Majorana fermions across the junction. We should expect that behavior of the persistent current differs in the cases of the strong and weak tunneling of Majorana fermions.

\section*{Model Hamiltonian, edge modes}

We consider a junction between two boundaries of the TSC. The lattice Hamiltonian for a $(p+ip)$-wave superconductor of spinless fermions consists of two terms: ${\cal H} = {\cal H}_{TSC} + {\cal H}_{tun}$. At that, the first term describes the TSC per se:
\begin{equation}
 {\cal H}_{TSC}= - \sum_{<ij>}a^\dagger_{i}a_j - 2\mu \sum_{j} n_j+
(i\Delta \sum_{<ij> x-links} a^\dagger_{i}a^\dagger_{j}+\Delta\sum_{<ij> y-links} a^\dagger_{i}a^\dagger_{j}+h.c.) ,
\label{eq-H}
\end{equation}
and the second term describes the tunneling of fermions between two boundaries of a TSC with a junction along the x-direction
\begin{equation}
{\cal H}_{tun}= - 2\tau e^{i\frac{Q}{2}} \sum_{x-links}  a^\dagger_ {x,1} a_{x,L} +h.c.,
\label{eq-Htun}
\end{equation}
where  $a^\dagger_{j} $ and $a_{j}$ are the spinless fermion operators on a site $j = {x,y}$ obeying usual anticommutation relations, and  $n_j$ denotes the density operator. The first term in (1) describes hoppings of spinless fermions between nearest-neighbor lattice sites with equal to the unity magnitude, $\mu$ is the chemical potential (by choosing $ 0 < \mu < 1$ we do not restrict the generality of the study). Remaining terms describe pairing with superconducting order parameter $\Delta > 0$, which is defined along the link. Links are divided into two types depending on their direction: real $\Delta$ along y-links and complex $i\Delta$ along x-links. In practice, values of $\Delta,|\mu| << 1$. Therefore, we consider low energy excitations for $\Delta,|\mu| < 1$. The term ${\cal H}_{tun}$ contains the tunneling amplitude $0 < \tau < 1$ and takes into account the applied flux $Q$. The value of $ Q$ is measured in units of the quantum of flux  $hc/(2e)$.

\begin{figure}[tp]
    \centering{\leavevmode}
\begin{minipage}[h]{1\linewidth}
\center{
           \includegraphics[width=\linewidth]{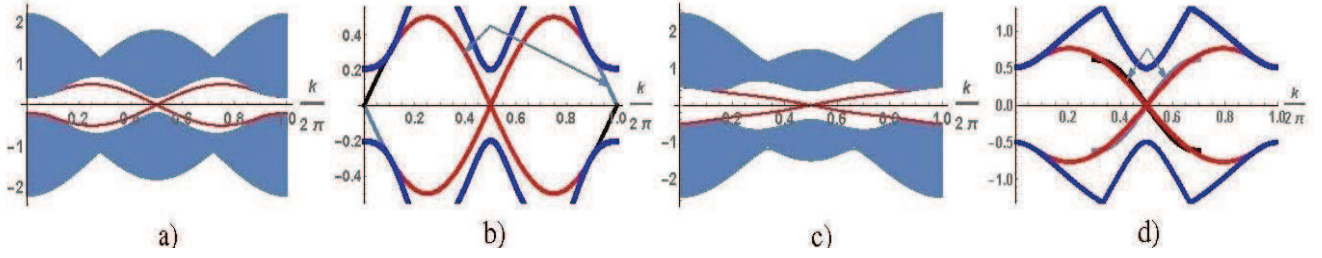}
                  }
    \end{minipage}
\caption{(Color online)  The band structure of TSC in the stripe geometry.
The bulk spectrum (denoted by blue lines) with edge modes (red lines) as a function of the momentum $k$ directed along the edge a),c), the low-energy modes of the bulk spectrum, edge modes and solution (\ref{A7}) for comparison (black lines also marked by arrows) b),d). The energies are calculated at $\Delta =\frac{1}{2}$,  $\mu=\frac{1}{5}$, $\tau =0$ a),b);
$\Delta =1$,  $\mu =\frac{1}{2}$, $\tau =\frac{1}{2}$, $Q=\pi$  c),d).
At $\mu\neq 0$  $\tau=0$ a) and $Q=\pi$,  $\tau<\tau_c$  c), the edge modes populate the gap, merge with the bulk states, intersect at the Dirac point $k = \pi$.
       }
    \label{fig:1}
\end{figure}
\begin{figure}[tp]
    \centering{\leavevmode}
    \begin{minipage}[h]{.5\linewidth}
\center{
    \includegraphics[width=\linewidth]{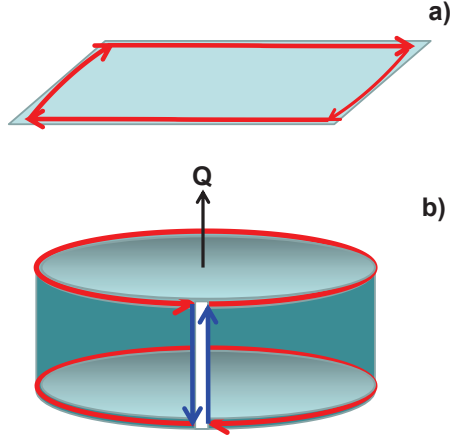}
                  }
    \end{minipage}
\caption{(Color online)
 Sketch of the superconductor in the form of a strip on the plane a), the superconductor has the spatial form of a hollow cylinder, and is pierced by a magnetic flux $Q$ b). Arrows indicate the direction of chiral current along the boundaries (red, blue).
  }
    \label{fig:2}
\end{figure}

Energies of spinless fermions E in the TSC that is described by the Hamiltonian (\ref{eq-H}) are arranged symmetrically with respect to the zero energy and are given by the following dispersion relation
\begin{equation}
E=\pm[(\mu+\cos k_x + \cos k_y)^2 +\Delta^2 (\sin^2 k_x +\sin^2 k_y)]^{1/2},
\label{eq-3}
\end{equation}
where the wave vector $\textbf{k}=\{k_x,k_y\}$. In a finite system, the one-particle spectrum of the Hamiltonian
 $ {\cal H}$ (\ref {eq-H}), (\ref {eq-Htun}), is also symmetric edge states including. The corresponding edge states are determined by the particle-hole states of Majorana fermions.

\begin{figure}[tp]
   \centering{\leavevmode}
          \includegraphics[width=1\linewidth]{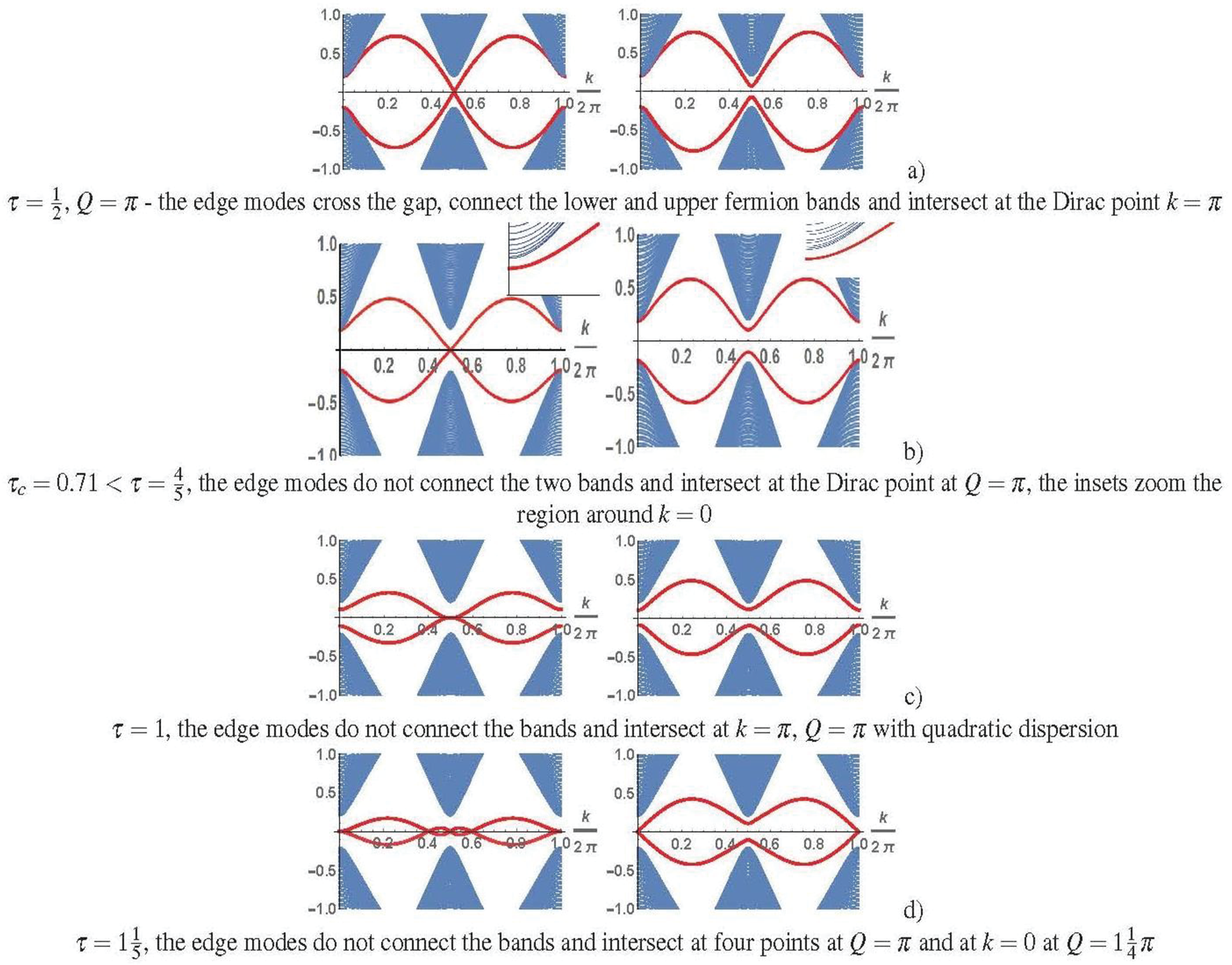}
\caption{(Color online) Low energy spectra with edge modes of the one-dimensional strip  along the \emph{x}-direction as a function of the momentum directed along the edge.
The energies are calculated at the Kitaev point $\Delta =1$ for $\mu=\frac{1}{5}$,  $Q=\pi$ left), $Q=1\frac{1}{4}\pi$ right) and for different $\tau$.
       }
    \label{fig:3}
\end{figure}

We analyze the formation of Majorana modes at the edges of the TSC. The gapped spectrum of excitations  (\ref{eq-3}) is realized in the topological nontrivial phase at $0<| \mu |<2$  (see the excitation spectra in Figs \ref{fig:1}a),c)). The topological properties of a system are manifested in the existence of a nontrivial Chern number $C$ and chiral gapless edge modes (see in Figs \ref{fig:1}), which are robust to effects of disorder and interactions. The excitation spectrum of the TSC includes chiral edge modes that connect the lower and upper fermion subbands. They are localized near the boundaries of the sample, and, therefore, amplitudes of the corresponding wave functions decrease exponentially with receding from the boundaries. The chiral gapless edge modes do exist in the gap if the Chern number of isolated bands located below the gap is nonzero. The gap of the superconductor collapses at $\mu=0$ and $\mu=\pm 2$. The TSC state with $C=\texttt{sgn} (\mu) $ is realized at  $|\mu|<2$ , whereas the Chern number is equal to zero in a trivial topological state at $|\mu| > 2$ \cite{D7}.

\begin{figure}[tp]
    \centering{\leavevmode}
    \begin{minipage}[h]{.6\linewidth}
\center{
       \includegraphics[width=\linewidth]{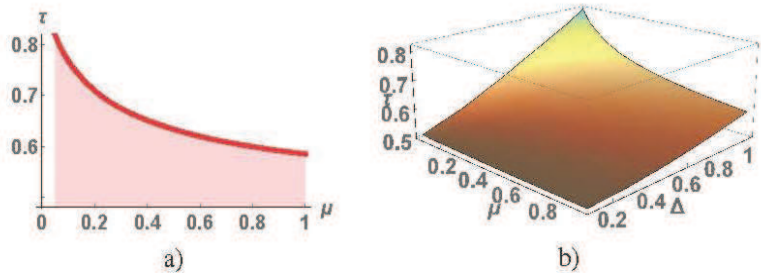}                   }
    \end{minipage}
\caption{(Color online) Numerical calculation of $\tau_c$ at $Q=\pi$ as a function of $\mu$,  $\Delta=1$ a) and
$\tau_c$ as a function of $\mu$ and $\Delta$ b).
       }
    \label{fig:4}
\end{figure}

\subsubsection*{The Kitaev point $\Delta=1$}

In order to describe in detail edge states of TSCs, we consider a superconductor in the form of a right square prism. Its base is taken to be LxL in size, while its height is assumed to be smaller than the superconducting coherence length. The superconductor can effectively be described by the 2D model (see Fig.\ref{fig:2}a)). In the Appendix, we rigorously prove that chiral edge modes exist in the TSC for open boundary conditions. At that, their energy is determined by the wave vector component $k$ that is parallel to the boundary. These edge modes intersect each other at the Dirac point according to the dispersion relation $E_{edge}=\pm \Delta \sin k $. This dispersion relation is valid up to the points, at which edge modes are entering the domain of bulk states (see Figs \ref{fig:1}). In the topological state with the Chern number equal to 1, gapless edge modes with wavevectors directed along the x-,y-boundaries have the corresponding Dirac points at $k_x = \pi , k_y =0$  and $k_x=0 , k_y=\pi$, respectively. In the topological state with $C=-1$  gapless edge modes have a different chirality, and their Dirac points are shifted to $\pi$ at $k_x=0$, $k_y =\pi$  and $k_x=\pi , k_y=0$, the TSC state is characterized by a chiral current along the boundary of the 2D system (see in Fig \ref{fig:2}a)). This current can be of different chirality depending on the sign of the Chern number.

The behavior of chiral edge modes at the junction (see in Fig.\ref{fig:2}b)) is examined for the sample in the form of a hollow cylinder with varying the applied flux $Q$. In the case of a contact interaction between fermions at the boundaries, the tunneling Hamiltonian can be expressed in Majorana operators $f_{x,1},g_{x,1}$ and $ f_{x,L},g_{x,L}$  as follows:  ${\cal H}_{tun}=- i \tau \cos(Q/2)\sum_{x-links} \delta_{1,L} (f_{x,1} g_{x,L} - g_{x,1} f_{x,L})$.
Gapless edge modes are associated with Majorana operators $g_1(k_x)$, $f_1(k_x)$ and $g_L(k_x)$, $f_L(k_x)$ that belong to the boundaries. Cases of $Q=\pm \pi$ are particular because the contact interaction between particles vanishes at the boundaries for arbitrary $\tau$. Thus, the system is reduced to the TSC with open boundary conditions, in which (as noted above) the chiral gapless modes are realized in the topological phase (see Figs \ref{fig:1}a),b)).At $\tau \neq 0$ and $Q\neq \pm \pi$ the edge modes at the junction are gapped, as a result of their hybridization at the Dirac point. In addition, we will demonstrate that their behavior depends on the magnitude of $\tau$. Majorana edge states are gapless at the points  $Q = \pm \pi$  with the linear dispersion in $k$:  $E_{edge}(\delta k)\sim \delta k$, $k=\pi+\delta k$ which is given by Eq (\ref{A7}) (see in Figs \ref{fig:1},\ref{fig:3}, the appendix contains some calculation details).

Numerical calculations show that in a weak tunneling regime $\tau<\tau_c$  two chiral gapless edge modes are realized in the spectrum of the TSC  for  $Q=\pm \pi$ (see in Figs \ref{fig:3}). These edge modes merge with the bulk states for any other $\tau$ and $Q$ in a weak tunneling regime. The edge modes with different chirality are localized at the different boundaries (at y = 1 and y = L) of the junction. The chiral edge modes yield chiral currents along the boundaries of the junction and form a chiral boundary current. The numerical calculation of $\tau_c$  at $Q=\pi$ as a function of $\mu$ is shown in Fig.\ref{fig:4}a) at the Kitaev point. The calculations of $\tau_c$  for arbitrary $Q$ demonstrate that $\tau_c$  has the maximum value at $Q=\pi$.

At the point $Q=\pi$ for $\tau >\tau_c$ gapless modes are localized at the junction, but they are non-chiral and do not touch fermion subbands at the arbitrary $Q$ (see Figs \ref{fig:3}b),c),d)). For $\tau =1$  the linear dispersion of edge modes vanishes at $k=\pi$. We see that the behavior of the edge modes changes radically at  $\tau >1$.  In the strong tunneling regime $\tau>\tau_c$ the edge modes are localized at both ($1$ and $L$) boundaries of the junction. They do not connect lower and upper subbands of the superconductor and form localized standing waves.

\subsection*{Arbitrary $\Delta$}

\begin{figure}[tp]
    \centering{\leavevmode}
      \begin{minipage}[h]{.4\linewidth}
\center{
    \includegraphics[width=\linewidth]{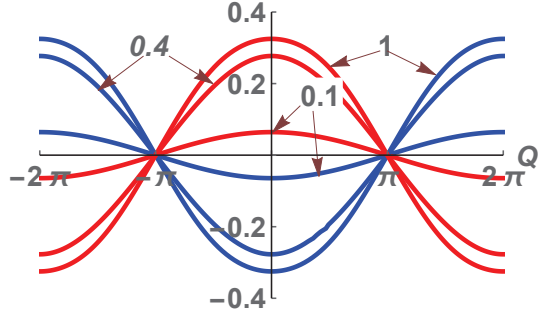}
                  }
    \end{minipage}
\caption{(Color online)
The gaps of edge modes with f (blue), h (red) fermion parity as functions of the flux Q calculated  at $\tau=0.1,0.4,1$, $\mu= \frac{1}{5}$, $\Delta =\frac{1}{2}$ .
       }
    \label{fig:5}
\end{figure}

The critical value $\tau_c$ depends on $\Delta$ and $\mu$. The minimal value of $\tau_c =\frac{1}{2}$ is reached in the $\Delta\rightarrow 0$ limit. The value  of $\tau_c$ calculated for $Q=\pi$ as a function of $\mu$ and $\Delta$ is shown in Fig.\ref{fig:4} b).
We have  plotted the width of the gap in the spectrum of Majorana bound states as a function of $Q$ for different values of $\tau$ (see in Fig\ref{fig:5}). It follows from numerical calculations that this  gap width is an even functions of $Q$, which can be
approximated by  $ \pm \tau^* \cos (Q/2)$, where, in the case of a weak tunneling the amplitude $\tau^*\sim \tau$ at $\tau <0.3$.

\section*{Persistent current}

The current along the boundaries is divided into chiral currents at the ends of the cylinder (red lines in Fig.\ref{fig:2}b)) and chiral currents along the junction (blue lines). Chiral currents at the ends of the cylinder are described by the Hamiltonian (\ref{eq-H}) with open boundary conditions and do not depend on the tunneling term (\ref{eq-Htun}), whereas currents along the junction are described by the total Hamiltonian $\cal H$.
The energy of the system $E^P(\tau,Q)=E_{cyl}^{p'} + E_{bulk}^p(\tau,Q)$ is determined by two terms: the energy of chiral edge modes at the ends of the cylinder with fermion parity $p'$ $E_{cyl}^{p'}$ (which does not dependent on $\tau$) and the energy of the superconductor, which takes into account the tunneling of fermions across the junction $E_{bulk}^p(\tau,Q)$, where $p,p'=f,h$ denote the fermion parity of the edge states: fermion (f) or hole (h),  the symbol $P$ denotes the fermion parity of the ground state.

\begin{figure}[tp]
    \centering{\leavevmode}
     \begin{minipage}[h]{.7\linewidth}
\center{
    \includegraphics[width=\linewidth]{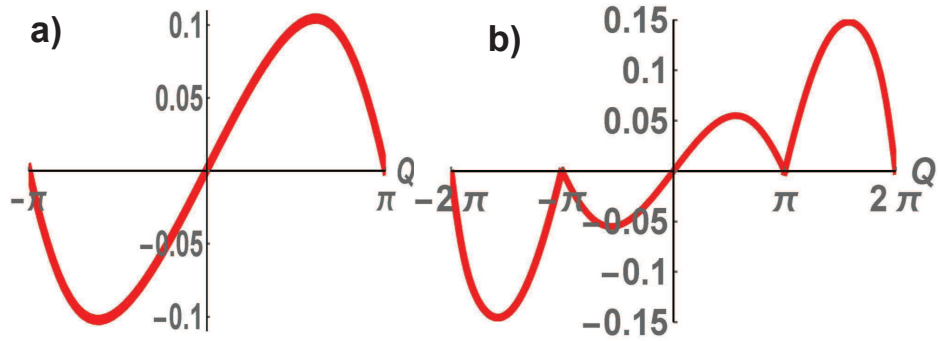}
                  }
    \end{minipage}
\caption{(Color online)
Persistent currents as function of the flux Q  at $\mu= \frac{1}{5}$, $\Delta =\frac{1}{2}$.  A $2\pi$-periodic persistent current in the regime of strong tunneling $\tau=\frac{7}{10}>\tau_c=0.57$ a),
a $4\pi$-periodic persistent current in the regime of weak tunneling $\tau=\frac{1}{2}<\tau_c=0.57$ b).
       }
    \label{fig:6}
\end{figure}

In the strong tunneling regime, the edge modes, which occur at the junction, are represented by localized standing waves at all Q's including the points $Q = \pm \pi$. Chiral currents at the ends of the cylinder and the current flowing along the junction, which is equal to zero, are not connected. Their fermion parities $p'$ and $p$ are not conserved. The fermion parities of the edge states are independent. In a contrast, in the weak tunneling regime, chiral currents flowing at the ends of the cylinder and along the junction are connected with each other due to the chiral current along the boundaries. Therefore, the fermion parities of Majorana-bound states located at the ends of the cylinder and at the junction are the same  $p=p'$.

Let us consider the behavior of the persistent current in the TSC in detail. In the limit $T\to 0$, the magnitude of the persistent current $I(\tau,Q)$  is determined by the ground-state energy of the system
$I(\tau, Q)=\partial E^P(\tau, Q)/\partial Q$ (in unities of $2e/\hbar$), where $E^P(\tau,Q)=\int dk \sum_{\epsilon_n (k) <0} \epsilon_n(k)$ is determined by the quasi-particle excitations $\epsilon_n (k)$, and  the Fermi energy is equal to zero at half-filling.
In the strong coupling regime of tunneling  $\tau>\tau_c$, the magnitude $I(\tau, Q)$ is a generic periodic function of the magnetic flux with the period of $2\pi$, so that  $I(\tau,Q)=\partial E^f(\tau, Q)/\partial Q$ (see in Fig\ref{fig:6} a)). In this case, the persistent current is determined by the energy of the superconductor, which takes into account both bulk excitations renormalized via the tunneling across the junction and the energies of edge modes at the junction. The fermion parity of edge states of Majorana fermions at the ends of the cylinder $p' = f$ and the Chern number, associated with these edge modes, are conserved. The fermion parity of edge Majorana fermions at the junction is not conserved. The system relaxes to the phase state with the minimum energy.

In strong coupling  tunneling regime  $\tau>\tau_c$,  $I(\tau,Q)$ is a typically periodic function of a magnetic flux with the period $2\pi$ (see in Fig\ref{fig:6} a)).
The persistent current is determined by the energy of the superconductor which takes into account the bulk excitations renormalized via the tunneling across the junction and the energies of edge modes at the junction. The fermion parity of edge states of Majorana fermions at the ends of the cylinder  $p`=f$  and the Chern number, associated with these edge modes, are conserved. The fermion parity of edge Majorana fermions at the junction is not conserved, the system relaxes to the phase state with the minimum energy.

In the weak tunneling regime $\tau<\tau_c$ all edge modes have the same chirality and fermion parity. This leads to a periodic persistent current having the period of  $4\pi$.. The fermion parities of edge modes, which form the current along the boundaries, are identical. At $Q<\pi$  and $Q>\pi$ the ground state energy is determined by $E_{cyl}^{p'}+E_{bulk}^p(\tau,Q)$ with $p'=p=f$.  At $Q>\pi$  the energy of the edge modes at the ends of the cylinder is negative, while the energy of the edge modes at the junction is positive. The balance of these energies determines the total energy of the system for given values of $\tau$ and $Q>\pi$. According to numerical calculations, a critical value of $Q_c$, at which energy difference of edge modes with different fermion parity changes its sign, is greater than $\pi$. The Chern number of the TSC is conserved, while the phase state of the system may not have the minimum energy at $Q>Q_c$. The resistive current is a periodic function of $Q$ with period $4\pi$, and $I(\tau,Q)$ is a continuous function of $Q$ within the whole interval $[-2\pi,2\pi]$ (see in Fig. \ref{fig:6} b)).

\section*{Conclusions}

This work is a step in our understanding of the behavior of a persistent current in topological systems.
We have discussed the emergence of a persistent current in 2D TSC, pierced by a magnetic flux. It is
proved that, the behavior of a persistent current is different in the case of strong and weak tunneling of
Majorana fermions across a junction. The fermion parity of edge modes, forming a current along the
boundaries of the sample, is the same, therefore the Chern number conserves a fermion parity of edge
modes in the case of a weak interaction. Bulk edge correspondence leads to $4\pi$-periodic tunnel current. In
a strong tunneling regime the currents at the ends of the cylinder and along the junction are not connected,
therefore the fermion parities of the edge modes at the ends of the cylinder and at the junction are not
conserved. At $Q=\pi$ in a strong tunneling regime spontaneous breaking of a fermion parity is realized. In
the absence of fermion parity conservation the system relaxes to the minimum energy state, thus triggering
a $2\pi$ periodic persistent current in TSC at strong tunneling of Majorana fermions across the junction.
The results can be generalized to other topological phases, in particular, to topological insulators.

\section*{Methods}
\subsection*{Edge modes in the 2D topological superconductor}

Below we discuss the solution of the Schr$\ddot{o}$dinger equation for the chosen Hamiltonian  ${\cal H}$  at the special point $\Delta=\pm 1$ using the formalism proposed for the calculation of Kitaev's chain  in Refs \cite{A1,A2}. We focus on a 2D superconductor in the form of a square with the LxL size. Its sketch is shown in Figs \ref{fig:2}. The wave function $\psi =\sum_{j=1}^L\sum_{s=1}^L[a^\dagger_{j,s}u_ {j, s}+a_{j,s}v_ {j,s}]$   is determined by amplitudes  $u_ {j,s}$ and $v_ {j,s}$,that are solutions of the following equations:
for $-L<j,s<L$
\begin{eqnarray}
&&(E +\mu) u_ {j, s} = -\frac{1}{2}( u_ {j + 1, s} + u_ {j - 1, s} +
      u_ {j, s + 1} + u_ {j, s - 1}) +
     \frac{i}{2} (v_ {j + 1, s} - v_ {j - 1, s}) +
    \frac{1}{2} (v_ {j, s + 1} - v_ {j, s - 1}),\nonumber\\
  &&  (E -\mu) v_ {j, s} =
    \frac{1}{2}( v_ {j + 1, s} + v_ {j - 1, s} + v_ {j, s + 1} +
      v_ {j, s - 1}) + \frac{i}{2} (u_ {j + 1, s} - u_ {j - 1, s}) -
      \frac{1}{2} (u_ {j, s + 1} - u_ {j, s - 1}),
        \label{A1}
\end{eqnarray}
for $s=1,L$, where $1<j<L$
\begin{eqnarray}
&&(E + \mu) u_ {j, 1} = -\frac{1}{2} (u_ {j,2} + u_ {j+1,1}+u_ {j-1,1} - v_ {j,2}) +
\frac{i}{2} (v_ {j+1, 1} - v_ {j-1, 1}) - \tau e^{-i\frac{Q}{2}} u_ {j,L},\nonumber\\
&&(E - \mu) v_ {j, 1} =\frac{1}{2}( v_ {j,2} + v_ {j+1,1} +v_ {j-1,1} -u_ {j,2}) +
\frac{i}{2} (u_ {j+1,1} - u_ {j-1,1}) +
   \tau e^{i\frac{Q}{2}} v_ {j,L},\nonumber\\
&&   (E +\mu) u_ {j,L} = -\frac{1}{2}(u_ {j,L - 1} +u_ {j+1,L}+u_ {j-1,L}+ v_ {j,L - 1}) +
 \frac{i}{2} (v_ {j+1,L} - v_ {j-1,L})- \tau e^{i\frac{Q}{2}}u_{j,1}, \nonumber\\
&&(E - \mu) v_ {j,L} = \frac{1}{2} (v_ {j,L - 1} + v_ {j+1,L} +v_ {j-1,L}+ u_ {j,L - 1}) +\frac{i}{2} (u_ {j+1,L} - u_ {j-1,L})+ \tau e^{-i\frac{Q}{2}}v_{j,1},
  \label{A2}
\end{eqnarray}
for $j=1,L$, where $1<s<L$
\begin{eqnarray}
(E + \mu) u_ {1, s} = -\frac{1}{2} (u_ {2,s} + u_ {1,s+1}+u_ {1,s-1}- v_ {1,s+1}+v_{1,s-1}) + \frac{i}{2} v_ {2,s},\nonumber\\
(E - \mu) v_ {1,s} =\frac{1}{2}( v_ {2,s} + v_ {1,s+1} +v_ {1,s-1} -u_ {1,s+1}+u_{1,s-1})+\frac{i}{2} u_ {2,s},\nonumber\\
(E +\mu) u_ {L,s} = -\frac{1}{2}(u_ {L-1,s} +u_ {L,s+1}+u_ {L,s-1} -v_ {L,s+1}+v_ {L,s-1})- \frac{i}{2}v_ {L-1,s}\nonumber,  \\
(E - \mu) v_ {L,s} = \frac{1}{2} (v_ {L - 1,s} + v_ {L,s+1} +v_ {L,s-1}-  u_ {L,s+1}+ u_ {L,s-1})
 - \frac{i}{2} u_ {L-1,s},
  \label{A3}
\end{eqnarray}
and the similar equations for the vertices of the square $\{1,1\};\{1,L\};\{L,1\};\{L,L\}$.

The solutions of Eqs (\ref{A1}) also satisfy  Eqs (\ref{A2})-(\ref{A3}) at $\tau =0$ and the following boundary conditions   $v_ {j,0} + u_ {j,0} = 0$, $v_ {j,L + 1} - u_ {j,L + 1} = 0$ and    $u_ {0,s} + i v_ {0,s} = 0$, $u_ {L + 1,s} - iv_ {L + 1,s} = 0$.
We determine the amplitudes of the wave function in accordance with the following Ansatz
\begin{multline}
u_{j,s}=A_u(k_x,k_y)e^{ik_x j+i k_y s}+B_u(k_x,k_y)e^{i k_x j-i k_y s}+C_u(k_x,k_y)e^{-i k_x j +i k_y s}+D_u(k_x,k_y)  e^{-i k_x j -i k_y s},\\
v_{j,s}=A_v(k_x,k_y)e^{i k_x j+i k_y s}+B_v(k_x,k_y)e^{i k_x j-i k_y s}+C_v(k_x,k_y)e^{-i k_x j +i k_y s}+D_v(k_x,k_y)  e^{-i k_x j -i k_y s}.
\label{A5}
\end{multline}
Unknown amplitudes in (\ref{A5}) are defined as $A_u(k_x,k_y)=G_u(k_x,k_y)$, $B_u(k_x,k_y)=G_u(k_x,k_y)e^{i(-\chi+ \alpha)}$, $C_u=G_u(k_x,k_y)e^{i(- \chi+ \beta)}$, $D_u(k_x,k_y)=G_u(k_x,k_y)e^{i\gamma}$,
$A_v(k_x,k_y)=G_v(k_x,k_y)$, $B_v(k_x,k_y)=-G_v(k_x,k_y)e^{i( \chi+ \alpha)}$, $D_v=G_v(k_x,k_y)e^{i(\chi + \beta)}$, $D_v(k_x,k_y)=-G_v(k_x,k_y)e^{i \gamma}$, where
$e^{2 i \chi} =\frac{i\sin k_y -\sin k_x}{i\sin k_y +\sin k_x}$,
the energies of the eigenstates are determined by Eq (\ref{eq-3}) at $\Delta=1$ and the constants $\alpha,\beta,\gamma$ are determined by the boundary conditions.
We redefine the  unknown $G_u(k_x,k_y)$ and $G_v (k_x,k_y)$ as
$G_u(k_x,k_y) =G \cos\varphi/2$ and $G_v(k_x,k_y) = iG \sin\varphi/2$, where $\tan \varphi =\frac{ \sin k_y - i \sin k_x}{\mu +\cos k_y  + \cos k_x}$, $G$ is a normalization constant.

Let us consider the points $k_y=0$ and $k_y=\pi$ at $\tau =0$ that correspond to zero energy of Majorana modes localized at the boundaries (see Figs \ref{fig:1} for the illustration). The solutions for particle-hole excitations localized at the boundary are determined by complex $k_y$-wave vectors $k_y=\pm i \varepsilon $ or $k_y=\pi \pm i \epsilon $ with
\begin{eqnarray}
 \varepsilon  =2\sinh^{-1}\left(\frac{1}{2}\sqrt{\frac{ \mu^2+4(1+\mu)\cos^2(k_x/2)-E^2}{-\cos k_x- \mu}}\right),\nonumber\\
- \cos k_x >\mu \nonumber\\
 \varepsilon  =2\sinh^{-1}\left(\frac{1}{2}\sqrt{\frac{ \mu^2+4(1-\mu)\sin^2(k_x/2)-E^2}{\cos k_x + \mu}}\right),\nonumber\\
 \cos k_x >-\mu
 \label{A6}
\end{eqnarray}
$\mu$ defines the bulk gap.
Solution (\ref{A6}) determines the momentum of an excitation at a given energy, we can invert Eq (\ref{eq-3}) yielding the momentum with energy E.
At $k_y=0,\pi$ or $k_x=0,\pi$  $\chi=\frac{\pi}{2}$ or $\chi=0$, therefore the boundary conditions  are reduced to the following equations $\sin[k_y(L+1)- \varphi] = 0$, $ \sin[k_x(L+1)]=0$.
Similar to the 1D model \cite{A1},  the energy of level localized at a boundary is equal to zero at $k_y= i \varepsilon , k_x=\pi$, in the $L\to \infty$ limit $E\sim (-1+\cosh\varepsilon -\mu)\exp(-2\varepsilon L)$. Complex solution for $k_y$ describes the edge modes localized at the $x$-boundary with $k_x$-dispersion. The boundary conditions describe free fermion states with the wave vector directed along the boundary. The solution of Eqs (\ref{A1}) are a x-y symmetric.

Let us consider the edge modes localized at the $x$-boundary with $k_y=i\varepsilon$ which have zero energy $E\to 0$ at the Dirac point $k_x=\pi$. The solution $E=0$ corresponds to the degenerate solution of Eqs (\ref{A1}) for the amplitudes of the wave function $u(k_x,i\varepsilon)=v(k_x,i\varepsilon)$. This solution is valid for arbitrary $k_x$ at $E^\ast\rightarrow E-\sin k_x=0$. We do not use the boundary conditions for calculation of the wave function, as a result, the dispersion of edge modes is determined for an arbitrary value of $\Delta$. We find that the dispersion relation for the energy of the edge modes reads: $E_{edge}(k_x)=\pm \Delta \sin k_x$. The numerical calculations of the spectrum of the edge modes, obtained for arbitrary $\mu$ and $\Delta$, confirm the dispersion (see in Fig.\ref{fig:1}b) for example). The energy of the edge modes at the $x-$ and $y-$boundaries have the intebtical dispersion for the wave vector directed along the boundary, that triggers a chiral current along the boundaries of the sample.

As we already noted above, points  $Q=\pm \pi$ are the special since the gapless edge modes are realized at them for $\tau \neq 0$. The zero-energy solutions for the edge modes at the Dirac point follow from the solutions of Eqs (\ref{A1})-(\ref{A3}) in the $L\to \infty$ limit at $\tau \neq 0$. Using appropriate boundary conditions we calculate a low energy dispersion of gapless edge modes at $Q=\pm \pi$.
The energies of the edge modes propagating along the junction have the following form
\begin{equation}
E_{edge}(\delta k_x)=\pm \frac{1}{3}\sin \delta k_x \mp\frac{2}{3} \sqrt{\sin^2 \delta k_x +3 w(0)} \cos (\zeta - 2 \pi/3)
 \label{A7}
\end{equation}
where $k_x=\pi+\delta k_x$, $w(Q)=(\mu+1-\cos \delta k_x)^2+\sin^2 \delta k_x + \tau^2  \cos Q$,
$\zeta =\frac{1}{3}\arccos\left(\frac{27}{54}\frac{-2 \sin^3 \delta k_x - 9  w(0) \sin \delta k_x + 27 w(Q)}{ (\sin^2 \delta k_x +3 w(0))^{3/2}}\right)$. Equation (\ref{A7}) is derived from the low energy solution of the following equation
$ E^3 - E^2 \sin \delta k_x- w(0) E +\sin \delta k_x w(Q)=0$.

We consider the zero $\delta k_x$ limit of Eq (\ref{A7}) and obtain the linear dispersion of edge modes at the Dirac point $E_{edge}(\delta k_x)=\pm v_{edge} \delta k_x$, where $v_{edge}=1-\frac{2\tau^2 }{\mu^2+\tau^2 }$. The linear dispersion of the edge modes vanishes at $\tau=1$. The solutions (\ref{A5}) do not satisfy the boundary conditions at $\tau=1$ and, as a result, solution (\ref{A7}) does not hold. According to numerical calculations the edge modes have a parabolic dispersion (see in Fig. \ref{fig:3}c)).

\section*{Author contributions statement}

I.K. is an author of the manuscript

\section*{Additional information}

The author declares no competing financial interests.

\end{document}